 \documentclass[preprint,showpacs,preprintnumbers,amsmath,amssymb]{revtex4}

\usepackage{graphicx}
\usepackage{dcolumn}
\usepackage{bm}

\begin{document}

\title{Anisotropic magneto-thermal transport and Spin-Seebeck effect}
 \author{J.-E. Wegrowe} \email{jean-eric.wegrowe@polytechnique.fr} \author{H.-J. Drouhin} 
\affiliation{Ecole Polytechnique, LSI, CNRS and CEA/DSM/IRAMIS, Palaiseau F-91128, France}
\author{D. Lacour} 
\affiliation{Institut Jean Lamour, UMR CNRS 7198, Universit\'e H. Poincarr\'e, F -5406 Vandoeuvre les Nancy, France}

\date{\today}

\begin{abstract}
 The angular dependence of the thermal transport in insulating or conducting ferromagnets is derived on the basis of the Onsager reciprocity relations applied to a magnetic system. It is shown that the angular dependence of the temperature gradient takes the same form as that of the anisotropic magnetoresistance, including anomalous and planar Hall contributions. The measured thermocouple generated between the extremities of the non-magnetic electrode in thermal contact to the ferromagnet follows this same angular dependence. The sign and amplitude of the magneto-voltaic signal is controlled by the difference of the Seebeck coefficients of the thermocouple.
 \end{abstract}

\pacs{85.80.Lp, 85.75.-d, 72.20.Pa \hfill}

\maketitle



Exploiting the properties of heat currents in magnetic systems for data processing and storage is a decisive challenge for the future of electronic devices and sensors. Recent research efforts, in line with spintronics developments, focus more specifically on the properties of the spin degrees of freedom associated to the electric charges and heat currents \cite{SpinCal}. Important series of experiments - the so-called spin-pumping and spin-Seebeck measurements - have been performed in this context  \cite{Uchida,Jaworsky,Uchida2,Sharoni,Huang,RezendePRL,Rezende,Chumak,Costache,Sandweg,Uchida3}: a magneto-voltaic signal is observed at zero electric current on electrodes that are in thermal contact with a ferromagnet. Surprisingly, for these experiments, the ferromagnetic layer used is indifferently conductor, semiconductor or insulator. 

A predominant role of spin degrees of freedom has first been assumed in order to interpret the experiments (typically in terms of transfer of pure-spin-current through the interface and inverse-spin-Hall effects inside the electrode). However, some authors suggest that anomalous Nernst effect (i.e., a thermoelectric bulk effect that is not explicitly spin-dependent) may suffice to explain the observations \cite{Ohno,Zink,Chien,Back,Yin,Short_Cut}. The problem is then to understand how the non-ferromagnetic electrode could mimic the behavior of the underlaying insulating or conducting ferromagnet \cite{Kikkawa}.  

The interpretation proposed here is based on the anisotropic thermal properties of ferromagnets (see Fig.1). It is shown that under heat current injection, an anisotropic temperature gradient is generated in the ferromagnet and transmitted to the electrode due to the thermal contact. This temperature gradient induces a thermocouple between the extremities of the electrode, which can be measured by a voltmeter. The anisotropic thermal transport equations is derived on the basis of the Onsager reciprocity relations. The signal follows the same angular dependence as the anomalous Hall effect (AHE) and the planar Hall effect (PHE). 

In conducting ferromagnets, the thermoelectric counterparts of AHE and PHE are well known in terms of anomalous and planar Nernst effect (ANE and PNE), that can be measured with a transverse electrode (i.e. on a Hall-cross  device: see Fig.1). ANE, AHE and AMR have the same spin-orbit scattering origin \cite{AMR,Ky,Nagaosa} which is responsible for the anisotropy of the electric response. The anisotropy of the thermal transport has been measured recently by Kimling et al. \cite{Nielsch} in Ni nanowires. This effect is generalized below to thermal transport in both conductors or insulators. The anisotropy is imposed by the direction of the magnetization with respect to the current flowing in the thin layer, i.e. by a unit vector $\vec m$ defined by the polar angle $\theta$ and the azimuthal angle $ \varphi$.

\begin{figure}
\includegraphics[scale=0.6]{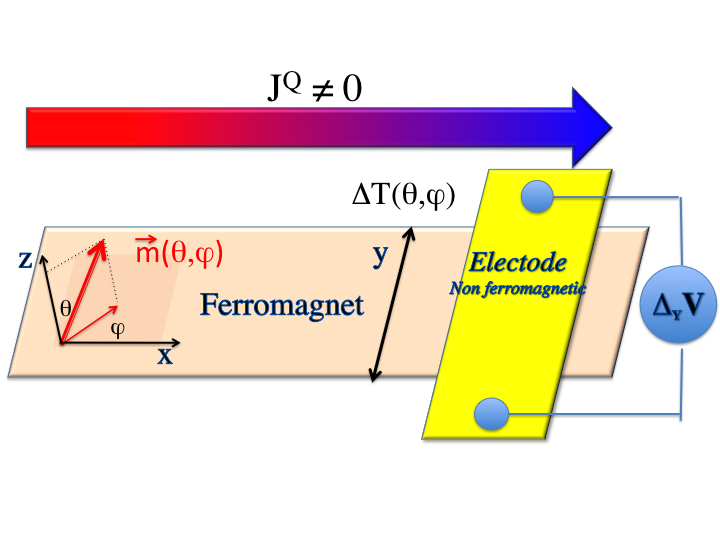}
\caption{Schematic view of Hall-cross device with a ferromagnet and a lateral non-ferromagnet electrode. The direction of the magnetization $\vec m$ is defined by the angles $\theta$ and $\varphi$. The heat current $\vec J^Q$ generates a temperature difference $\Delta T(\theta, \varphi)$ that is angular dependent. This temperature difference can be measured by the Seebeck voltage $\Delta_y V$.}
\label{fig1}
\end{figure}

\subsection{Electric transport equations}

The AMR effect is described by the Ohm's law, with the help of the Onsager reciprocity relations. An electric current $\vec J^e$ is defined locally (at each point of the material) with the help of a conductivity tensor $\hat \sigma$ such that 
$\vec J^e = -\hat \sigma \, \vec \nabla \mu^e/q$, 
 where $\mu^e$ is the local electrochemical potential and $q$ is the electric charge. It a thin ferromagnetic layer is oriented in the plan \{$\vec e_x$,$\vec e_y$\} {\it perpendicular to the orientation of the magnetization} (the general case of an arbitrary polarization orientation $(\theta, \varphi)$ is considered below), the ohm's law reads : 
\begin{equation}
\left( \begin{array}{c}
J^e_{x} \\
J^e_{y }  \\
J^e_{z } \\
\end{array} \right)
=  - \frac{1}{q} \left( \begin{array}{ccc}
                       \sigma & \sigma_{H} & 0 \\
                       - \sigma_{H}  &  \sigma & 0  \\
                        0 & 0 & \sigma_{z } \\
		\end{array} \right)
                        \left( \begin{array}{c}
\partial \mu^e/ \partial x \\
\partial \mu^e/\partial y\\
\partial \mu^e/ \partial z \\
 \end{array} \right).
 \label{Conductivity}
\end{equation}
 where the off-diagonal coefficients $\sigma_{H}$ is defined by the microscopic mechanisms that couples the spin-dependent electric carriers and the magnetization (skew scattering, side-jump scattering, intrinsic effect, etc) \cite{Nagaosa}. However, whatever the details of the microscopic origin, {\it the microscopic time invariance symmetry is associated to spin-rotational symmetry}, so that the Onsager reciprocity relation imposes \cite{OnsagerII,DeGroot} $\sigma_{xy} = - \sigma_{yx} \equiv \sigma_{H}$.  We assume furthermore that the material is isotropic at zero magnetization (most of the experiments are performed on polycrystalline materials). We have consequently $\sigma_{xx} = \sigma_{yy} = \sigma$, but $\sigma_{z} \ne \sigma$ due to the magnetization anisotropy. 
 

Since the experiments are usually performed in the galvanostatic mode (i.e., constant current $\vec J^e$), it is convenient to invert the Ohm's law Eq.(\ref{Conductivity}) with introducing the resistivity tensor $\hat \rho = \hat \sigma^{-1}$. In the vectorial form we have:
\begin{equation}
- \vec \nabla \mu^e/q = \rho \vec J^e + \left ( \rho_{z} - \rho \right ) \, \left (  \vec J^e. \, \vec m \right ) \vec m \, + \rho_{H} \,  \vec m \times \vec J^e
\label{AMR}
\end{equation}
where  $\rho = \sigma/(\sigma^2 + \sigma_{H}^2)$, $\rho_{H} =  \sigma_{H}/(\sigma^2 + \sigma_{H}^2)$ and $\rho_{z} =  \sigma_{z}^{-1}$. Eq.(\ref{AMR}) is the general vectorial expression for electronic transport that takes into account spin-orbit interactions \cite{AMR}. After integration over the sample, the first term in the right hand side of Eq.(\ref{AMR}) defines the usual resistance, the second term is the AMR  effect (including the planar Hall contribution), and the last term defines the anomalous Hall effect.

\subsection{Thermal transport equations}

In this section, we treat a system which is not in contact to an electric generator. Instead, the ferromagnetic degrees of freedom are excited by a specific energy source. This is the case for magnetic fields (e.g. ferromagnetic resonance excitation or magnetization reversal), temperature gradient, magneto-optic excitations, magneto-mechanic excitations, etc. We consider here only the situation for which the heat currents $ \vec J^Q$ is equal to the energy current $ \vec J^u$ (the thermodynamic relation $ \vec J^Q = \vec J^u - \tilde \mu^e \vec J^e$ links the two currents \cite{Wegrowe13}). 

In the case of electric conductors, the thermoelectric transport equations are well-known  \cite{DeGroot}. For electric current we have  
\begin{equation}
\vec J^e = - \frac{\hat \sigma}{q} \, \vec \nabla  \mu^e 
- \hat \sigma \hat{\mathcal S} \,  \vec \nabla T,
\label{ThermoCurrentA}
\end{equation}
and for heat current we have:
\begin{equation}
\vec J^{Q} =     -\hat \lambda \, \vec \nabla T - \hat \Pi \, \vec \nabla  \mu^e
\label{ThermoCurrentB}
\end{equation}
where $\hat \lambda $ is the heat conductivity tensor, $\hat{\mathcal S}$ is the Seebeck tensor, $\hat \Pi = -T \hat \sigma \hat{\mathcal S}$ is the Peltier tensor. 
 In the case of isotropic materials, all matrices verify the same Onsager symmetry relation and the matrices $\hat \lambda$ and $\hat{\mathcal S}$ have the same form as that of the electric conductivity Eq.(\ref{Conductivity}) \cite{DeGroot}:
\begin{equation}
\hat \lambda = 
\left( \begin{array}{ccc}
                      \lambda & \lambda_{RL} & 0 \\
                        - \lambda_{RL}  & \lambda & 0  \\
                        0 & 0 & \lambda_{z} \\
	\end{array} \right)
 \qquad
\hat{\mathcal S} = 
\left( \begin{array}{ccc}
                       \mathcal S & \mathcal S_{N} & 0 \\
                        -\mathcal S_{N}  & \mathcal S & 0  \\
                        0 & 0 & \mathcal S_{z} \\
	\end{array} \right),
 \label{Seebeck}
\end{equation}
where  $\lambda$ is the the thermal conductivity, $\lambda_{RL}$ is the Righi-Leduc coefficent,  and the anisotropy $\lambda_z \ne \lambda$ is attested by the thermal conductance measurements performed on ferromagnets \cite{Nielsch}$. Similarly, \mathcal S$ and $\mathcal S_z$ are the Seebeck coefficients and $\mathcal S_{N}$ is the Nernst coefficient.

In the case of ferromagnetic insulators, Eq. (\ref{ThermoCurrentA}) is zero, and Eq. (\ref{ThermoCurrentB}) is modified in order to introduce the relevant transport coefficients. The total chemical potential $\tilde \mu^{F}$, related to the ferromagnetic degrees of freedom, can be defined with the effective magnetic field: $\vec H_{eff} \equiv - \vec \nabla \tilde \mu^F$,  where $\tilde \mu^F$ contains a drift term and a diffusion term $ \tilde \mu^{F} = \mu^F + kT \, ln(n^F)$ (the drift ferromagnetic chemical potential $\mu^F$ and the density of magnons $n^F$  have been defined in references \cite{Wegrowe13,PRB08,Entropy} for the uniform mode). In the insulating ferromagnet, the anisotropic thermal transport equation is described by the Onsager matrix $ \hat{\mathcal L}$ \cite{Wegrowe13}:
\begin{equation}
\vec J^Q = \hat{\mathcal L} \, \vec \nabla \tilde \mu^F,
\label{ThermoCurrentC}
\end{equation} 
where the cross-coefficient of the matrix $\hat{\mathcal L}$ follows the same reciprocity relation $\mathcal L_{xy} = \mathcal L_{yx} = \mathcal L_{RL }$ as for the electric conductivity tensor $\hat \sigma$, the thermal conductivity tensor $\hat \lambda$ and the Seebeck tensor $\hat{\mathcal S}$, while the anisotropy $\mathcal L_z \ne \mathcal L$ is due to the anisotropy of the ferromagnet \cite{Nielsch,Wegrowe13}. Eq. (\ref{ThermoCurrentC}) reads:
\begin{equation}
 \left( \begin{array}{c}
	J^Q_{x } \\
	J^Q_{y }  \\
	J^Q_{z } \\
 \end{array} \right)
=  \left( \begin{array}{ccc}
                       \mathcal L & \mathcal L_{RL} & 0 \\
                        - \mathcal L_{RL}  &  \mathcal L & 0  \\
                        0 & 0 & \mathcal L_{z} \\
		\end{array} \right)
		\left( \begin{array}{c}
 \frac{\partial \tilde \mu^F}{\partial x} \\
\frac{\partial  \tilde \mu^F}{\partial y}\\
\frac{\partial \tilde \mu^F}{\partial z} \\
\end{array} \right)
\label{Thermal_AMR}
\end{equation}
or, in vectorial form:
\begin{equation}
\vec J^Q = \mathcal L \vec \nabla \tilde \mu^F + \Delta \mathcal L \, \left (  \vec \nabla \tilde \mu^F. \vec m\right ) \vec m - \mathcal L_{RL}  \, \vec m \times \vec \nabla \tilde \mu^F
\label{Thermal_AMR_B}
\end{equation}
where $\Delta \mathcal L =  \mathcal L_{z} - \mathcal L$ is the thermal anisotropy and $\mathcal L_{RL}$ is the Righi-Leduc coefficient \cite{DeGroot} (or ``magnon Hall'' coefficient for ferromagnetic insulators \cite{thermal_Hall1,thermal_Hall2}). Note that Eq.(\ref{Thermal_AMR_B}) is formally equivalent to the general expression for the AMR (Eq.(\ref{AMR})) if we replace the heat current $\vec J^Q$ by the electric current $\vec J^e$, the generalized force $\vec \nabla  \tilde \mu^F$ by the electric field $\vec E$, and the electric transport coefficients by the thermal transport coefficients:
\begin{equation}
\vec \nabla \mu^F = \frac{1}{( \mathcal L^2 +  {\mathcal L}_{RL}^2)} \left (\mathcal L \vec J^Q + \left ( \frac{\mathcal L^2 +  {\mathcal L}_{RL}^2}{{\mathcal L}_{z}} - {\mathcal L} \right ) \, \left (  \vec J^Q. \, \vec m \right ) \vec m \, + {\mathcal L}_{RL}  \, \vec m \times \vec J^Q \right )
\label{Thermal_AMR_C}
\end{equation}
Eq.(\ref{Thermal_AMR_C}) should be used if the stationary heat current $\vec J^Q$ is imposed by external sources while Eq.(\ref{Thermal_AMR_B}) is convenient if the generalized force $\vec \nabla \tilde \mu^F$ is imposed (this is the same distinction as for galvanostatic or potentiostatic configurations for electric measurements). The appropriate description depends on the experimental boundary conditions. In the context of spin pumping and Spin Seebeck experiments, the heat current is imposed by the boundary conditions, and the local gradient of the chemical potential (near the electrode) can be reduced to a gradient of temperature $\vec \nabla T$. Eq.(\ref{Thermal_AMR_C}) reads:
\begin{equation}
\vec \nabla T = r \vec J^Q + \Delta r \left (  \vec J^Q. \vec m \right ) \vec m \, + \tilde r_{RL} \, \vec m \times \vec J^Q
\label{Righi_Leduc}
\end{equation}

where $r = \lambda/(\lambda^2 + \lambda^2_{RL})$ and $r_{RL} = \lambda/(\lambda^2 + \lambda^2_{RL})$ are the thermal resistivity and $\Delta r = 1/\lambda_z- \lambda/(\lambda^2 + \lambda^2_{RL})$ is the anisotropic thermal resistivity.
 
\subsection{Anisotropic magneto-thermal transport}


In the previous discussion, the orientation $\vec m$ of the magnetization was taken perpendicular to the plane $(Ox,Oy)$ of the thin ferromagnetic layer, and the Hall and Nernst voltages were measured perpendicular to both the heat current and the magnetization. However, in the experimental context, the magnetization is oriented at an arbitrary direction with respect to the reference frame defined by the plane of the ferromagnetic layer $(\vec e_x, \vec e_y)$ and the unit vector $\vec e_z$ perpendicular to both $\vec e_x$ and $\vec e_y$ . 

 If the magnetization $\vec m$ is oriented at the polar angle $\theta$ and the azimuthal  angle $\varphi$ we have $\vec m = m_x \vec e_x + m_y \vec e_y + m_z \vec e_z$, where $m_x= sin(\theta) cos(\varphi)$, $ m_y = sin(\theta) sin(\varphi)$ and $m_z = cos(\theta)$. The angular dependence of the temperature gradient $\vec \nabla T(\theta,\varphi)$ generated by the anisotropic heat current $J^Q(\theta,\varphi)$ is given by Eq.(\ref{Righi_Leduc}): 

 \begin{equation}
 \begin{array}{c}
\vec \nabla T(\theta,\varphi)=\\
 \left(
\begin{array}{c}
\, \, \, \, \, \,  \left ( r  +   \Delta r \, m_x^2  \right ) J^Q_x \, \, \, \, \, \, + \left ( \Delta r \, m_x m_y  - r_{RL} m_z \right ) J^Q_y + \left (   \Delta r \, m_x m_z  + r_{RL} m_y \right ) J^Q_z
\\
 \left ( \Delta r \, m_y m_x  + r_{RL} m_z \right ) J^Q_x + \, \, \, \, \, \,  \left ( r  +   \Delta r \, m_y^2  \right ) J^Q_y \, \, \, \, \, \, + \left (   \Delta r \, m_x m_z  - r_{RL} m_x \right ) J^Q_z
\\
\left ( \Delta r \, m_z m_x  - r_{RL} m_y \right ) J^Q_x + \left ( \Delta r \, m_z m_y  + r_{RL} m_y \right ) J^Q_y + \, \, \, \, \, \,  \left (r  +   \Delta r \, m_z^2  \right ) J^Q_z \, \, \, \, \, \,
 \end{array}
\right)
\end{array}
\label{Righi_Leduc2}
\end{equation}

In well designed experimental situations, the heat current $\vec J^Q$ is oriented along the $\vec e_x$ direction only (in the case of a heat current oriented perpendicular to the plane $\vec J^Q = J^Q_z \vec e_z$ the result is the same with the transformation $\theta + \pi/2 \rightarrow \theta $). The temperature gradient generated in the ferromagnet is then given by: 
\begin{equation}
\vec \nabla T = \left (
\begin{array}{c}
 r \, +  \Delta r \, sin^2(\theta) cos^2(\varphi) \\
\Delta r \, sin^2(\theta) cos(\varphi) sin(\varphi) +  r_{RL} \, cos(\theta)\\
\Delta r \, cos(\theta) sin(\theta) cos(\varphi) - r_{RL } \, sin(\theta) sin (\varphi) 
\end{array}
\right) J^Q_x
\label{Righi_LeducB}
\end{equation}

\subsection{Thermocouple effect on the electrode} 
 
A temperature difference generated by $\vec \nabla T(\theta,\varphi)$ in the ferromagnet layer can be measured on an electrode thanks to the thermocouple effect. The electric potential $\Delta V_y = - \int_C^D \partial \mu^e/(q \partial y)dy$ generated along $\vec e_y$ is given by Eq.(\ref{ThermoCurrentA}) with $\vec J^e = 0 $: 
\begin{equation}
\Delta V_y = \Delta \mathcal S \, \Delta_y T + \Delta \mathcal S_{N} \, \Delta_x T
\label{Result}
\end{equation}
 where $\Delta_x T = \int_C^D \partial T/\partial x \, dy$ and $\Delta_y T = \int_C^D \partial T/\partial y \, dy$. $\Delta \mathcal S$ and $\Delta \mathcal S_N$ are the Seebeck and Nernst coefficients of the electrode (typically Pt) with respect to the Seebeck coefficients of the electric wires used to measure the potential (typically Cu, Ag or Au that have close Seebeck coefficients). In the expression of $\Delta V_y$ Eq. (\ref{Result}), the term proportional to the normal Nernst coefficient of the non-ferromagnetic electrode $\Delta {\mathcal S}_{N}$, has the same angular dependence as the AMR measured along the current direction: $\Delta_x T \propto r \, +  \Delta r \, sin^2(\theta) cos^2(\varphi)$. However, like for normal Hall effect in a moderate external magnet field ($< 10$ T), we have ${\mathcal S}_{N} \ll \mathcal S $, so that only the first term should be significant (see however the results at high fields reported in \cite{Shi}).
The principal contribution to the anisotropic magneto-thermal effect $\Delta_y V$ in Eq.(\ref{Result}) contains two terms, corresponding to "anomalous" and "planar" Hall or Nernst effects:
\begin{equation} 
\Delta V_y \approx
J^Q_x  \Delta \mathcal S  \left ( \Delta r sin^2(\theta) cos(\varphi) sin(\varphi) +  r_{RL} cos(\theta) \right)
\label{thermocouple}
\end{equation} 
The second term in the right hand side of Eq.(\ref{thermocouple}) (the ``anomalous" contribution), proportional to $cos(\theta)$ corresponds to the {\it magnetization-perpendicular-to-the-plane} (MPP)  geometry. The maximum signal (given by the difference between the voltages measured at $\theta = 0$ and $\theta = \pi/2$ for $\varphi = 0$) is $\Delta V_{y}^{MPP} = J^Q_x \Delta \mathcal S r_{RL}$. The first term in the right hand side of Eq.(\ref{thermocouple}) (the ``planar" contribution) corresponds to the {\it magnetization-in-the-plane} geometry (MIP)   (measured between $\varphi = 0$ and $\varphi = \pi/2$ for  $\theta = 0$) is $\Delta V_{y}^{MIP} = J^Q_x \Delta \mathcal S \Delta r$. 

In the case of MIP, the anisotropic magneto-thermal coefficient $\Delta r/r$ has been measured recently by Kimling et al. \cite{Nielsch} in Ni nanowires, and is of the same order than the AMR (i.e., one percent). For a typical Hall cross (the width $L$ of the ferromagnetic thin film is the same as the width of the transverse electrode), we have $\Delta V_{y}^{MIP} = J^Q \Delta \mathcal S (\Delta r/r)/(\lambda d)$, where $d$ is the thickness of the thin layer, which is typically $10$ $nm$.  This thickness dependence of the form $1/d$ is due to the simplest geometry for the thermal resistance ($r = L/(\lambda L d)$) and for the Seebeck thermocouple (constant $\Delta \mathcal S$ for identical section $Ld$ of the ferromagnetic layer and the electrode placed in continuity). The heat conductivity is of the order of $\lambda \approx 100$ $W/(mK)$, and the Seebeck coefficient difference  is of the order of $\Delta \mathcal S \approx 1$  $\mu V/K$ for Pt-noble metal contacts.  With $\Delta r/r \approx 1\%$ and heat power $J_x^Q \approx 1$ $mW$ we have $ \Delta V_y \approx 10$ $\mu V$.

Note that for both MIP and MPP signals, {\it the magneto-voltaic signal vanishes if the Seebeck coefficient of the electrode is equal to that of the wire used for the electric contacts}: $\Delta \mathcal S \approx 0 \mu V$ (typically for $Cu$, $Au$ and $Ag$ electrodes, as already observed), and the sign of $ \Delta V_y$ is inverted by choosing an electrode for which  $\Delta \mathcal S < 0$ (this is the case with using Ta instead of Pt).  


\subsection{Conclusion} 
The application of the Onsager reciprocity relation to heat transport coefficients in a ferromagnet leads to consider an anisotropic heat transport effect in ferromagnetic conductors and insulators (anomalous Righi-Leduc effect or thermal Hall effect). A characteristic angular dependence of the temperature gradient is predicted, that follows that of anomalous Nernst (magnetization perpendicular-to-the-plane) and planar Nernst (magnetization in-the-plane) effects. The amplitude of the first effect is proportional to the anisotropic thermal resistance $\Delta r$, and the second effect is proportional to the Righi-Leduc thermal resistance $r_{RL}$. The thermocouple effect generated by this temperature gradient on the non-magnetic electrode in thermal contact to the ferromagnet induces a voltage $ \Delta V_y = J_x^Q \Delta \mathcal S  \left ( \Delta r \, sin^2(\theta) cos(\varphi) sin(\varphi) +  r_{RL} \, cos(\theta) \right )$ that mirrors this specific angular dependence. 


 \end{document}